\documentclass[prb,aps,twocolumn]{revtex4}
\usepackage{bm}
\usepackage{graphicx}
\usepackage{amsfonts}
\usepackage{amsmath}
\usepackage{amssymb}
\newcommand {\be}{\begin{equation}}
\newcommand {\ee}{\end{equation}}
\newcommand {\bea}{\begin{eqnarray}}
\newcommand {\eea}{\end{eqnarray}}

\begin{document}


\date{\today}

\title{ The electron-hole superfluidity in two coaxial nanotubes}
\author{Oleg  L. Berman, Ilya Grigorenko,  Roman Ya. Kezerashvili}
\affiliation{Physics Department, New York City College of Technology, The City University of New York,
Brooklyn, NY 11201, USA}

\begin{abstract}The superfluid phase and Coulomb drag effect caused by the
pairing in the system of spatially separated electrons and holes in two coaxial cylindrical 
nanotubes are predicted. It is found that the drag resistance as a function of temperature experiences a 
jump at the critical temperature 
and can be used for the manifestation of the superfluid transition. 
It is demonstrated that at sufficiently 
low temperatures the order parameter and free energy density exhibit 
a kink due to the electron-hole asymmetry that is controlled by the radii of the nanotubes.


\end{abstract}

\maketitle

\section{Introduction}

\label{intro}

The superfluid state for a dense
system of spatially separated electrons and holes in two
parallel layers was predicted in Ref.~\onlinecite{Lozovik}, and physical properties of this state have
been studied \cite{Pogrebinskii,VM,BJL}.
The peculiarity of such systems is that
the pairing between the carriers occurs not via the usual weak  electron-phonon interaction
 mechanism like in  conventional superconductors,
but through the much stronger electron-hole Coulomb
attraction. When the strength of the mutual attraction in dense systems is
sufficiently weak, a standard Bardeen-Cooper-
Schrieffer (BCS) description~\cite{Schrieffer} is applicable.
Another limiting case is characterized by a sufficiently strong interaction  in
 dilute systems, when fermions form bounded
 pairs, which can be described as
composite bosons (indirect excitons), which undergo Bose-Einstein
condensation (BEC). In the both limits the Coulomb attraction between electrons and holes can 
introduce Coulomb drag that is a process in spatially separated conductors, which 
enables a current flowing in one of the conductors to induce a voltage drop in the other one.
In the case when the second conductor is a part of closed circuit, the induced current flows.
The experimental observation of the exciton
condensation and perfect Coulomb drag was claimed recently for the
spatially separated electrons and holes in GaAs/AlGaAs coupled
quantum wells at the presence of high magnetic field perpendicular
to the quantum wells~\cite{Nandi}. A steady transport current of
electrons driven through one quantum well was accompanied by an
equal current of holes in the other. The intermediate regime, the so
called BCS-BEC crossover, is an interesting phenomenon by itself
\cite{strinati}, but it is  beyond  the present study.
The BCS regime is achieved for relatively high fermionic densities
\cite{Lozovik} and it is a subject of the present work.

In Ref.~\onlinecite{Pogrebinskii} the
authors discussed the drag of holes by electrons in a
semiconductor-insulator-semiconductor structure. The prediction was
that for two conducting layers separated by an insulator there will
be a drag of carriers in one layer due to the direct Coulomb
attraction with the carriers in the other layer.
The Coulomb drag effect in
the electron-hole two-layer BCS system was also analyzed in
Ref.~\onlinecite{VM}. If the external potential difference is applied to one of
the layers, it will cause the electric current in the other layer. The
 current in another layer will be initiated due to the correlations between
electrons and holes at temperatures below the critical one. Let's mention also
that the theory of the drag effects in the system of spatially separated electrons
and excitons in an optical microcavity developed in Refs.~\onlinecite{BKL1,BKL2} predicts
that at low temperature an electron current induces the polariton flow, while
the electron current dragged by the polariton flow is strongly suppressed below polariton
superfluid transition temperature. This demonstrates the asymmetry of the drag processes in the system.
In all above mentioned studies possible asymmetry between electrons
and holes excitation spectra is caused by the difference between the effective
masses of the carriers. In this work we present the study of the asymmetry in 
the excitation spectra of electrons and holes and its effect on the Coulomb the drag process 
due to different radii of two coaxial nanotubes.

Let us consider a system of two coaxial cylindrical nanotubes separated by a
dielectric with the spatially separated electrons and holes confined
on  each nanotube, as shown in Fig. 1. We study the
formation of the superfluid phase resulting in the electron-hole
Coulomb drag effect in this system. The electrons on the outer
nanotube and holes on the inner one have different excitation
spectra due to the different radii of the nanotubes. As a
result, the  conductivities  for the inner and
outer nanotubes can substantially differ from each other and also
strongly depend on the radii of the nanotubes. By measuring  the
drag conductivity as a function of the temperature one can observe
the superfluid transition in the system, and by measuring the
jump in the drag coefficients one can obtain the critical
temperature of the BCS phase transition causing the superfluidity.
The paper is organized in the following way. In Sec.~\ref{geometry}
we discuss the particularities of  the physical properties
of the system originated from the cylindrical geometry of two
coaxial nanotubes. The BCS state for the spatially separated
electrons and holes in two coaxial nanotubes is described in
Sec.~\ref{bdg}. In Sec.~\ref{drag_coef} the calculation of the
transconductivity coefficients is presented. Finally, the discussion
of the results and
 conclusions follow in Sec.~\ref{disc}.

\section{Two coaxial  nanotubes}

\label{geometry} We consider a system of two coaxial cylindrical
nanotubes and assume that the inner nanotube is doped by
holes, while carriers on the outer nanotube are electrons.
The geometry of two coaxial nanotubes leads to significant
differences compared to a system of two parallel plane layers
\cite{Lozovik,VM,BJL},
 and promises much richer physics, compared to the plane geometry.
 Unlike  the case of two plane layers, the conductivities for the
inner and outer nonotubes can differ because of their different
curvatures. The quantum confinement in the coaxial nanotubes system
may result in enhancement of the order parameter for
smaller radii of the nanotubes. Similar confinement effects were
studied in thin superconducting layers~\cite{Shanenko} and in
metallic cylindrical superconducting nanowires~\cite{Croitoru_sup}.

\begin{figure} 
\includegraphics[width=8cm,angle=0]{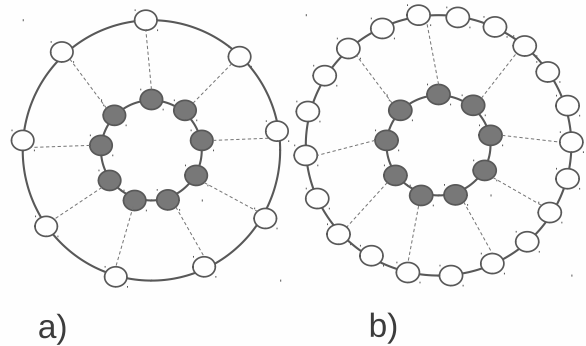}
\caption{{\protect\footnotesize {The distribution  of holes (dark circles) on the inner
nanotube and electrons (white circles) on the outer one. a) the case of equal amount of electrons and holes,
resulting in different carrier densities; b) the case of equal densities of holes and electrons.
Note that not all electrons on the outer nanotube can find a partner hole.
 }}}
\label{label_figure1}
\end{figure}

For two coaxial cylindrical nanotubes there are two interesting cases: i)
both nanotubes have equal  amount of electrons and holes; ii)
both nanotubes have equal densities of electrons and holes. If the
numbers of electrons and holes are equal, as shown in
Fig.~\ref{label_figure1}a, the carrier concentrations and the chemical
potentials for the electrons and holes on the outer and the inner
nanotubes will be different. One would expect to observe a lower
chemical potential in the outer nanotube. Even though at low
temperatures all electrons and holes can be paired,  it is expected that
the mismatch between the chemical potentials $\mu_e\ne\mu_h$ may
significantly reduce the critical temperature of the superconducting
transition. For sufficiently large mismatch between the chemical potentials, comparable to the order parameter,
the BCS transition may be impossible.
One has also to note that even in the case of perfectly
equal carrier concentrations $\rho_e=\rho_h$ the chemical potentials
may be slightly different due to the effect of the different
curvatures of the nanotubes. This difference is expected to be
noticeable for relatively small, up to several
nanometers, but different radii  of the nanotubes.

In the case of equal densities of the carriers on each nanotube,
the total number of the carriers on the outer
 nanotube will be higher then on the
inner one, as this is shown in Fig.~\ref{label_figure1}b.
At low temperatures this will result in a considerable amount of the
carriers on the outer nanotube (electrons), which can not find their
pairs on the inner nanotube (holes). In this case the conductivity
in the system will be defined by the paired and unpaired components
of the system. One has to note that we consider only the case of relatively weak
imbalance between the carriers. A larger difference  between  the numbers of the electrons and holes 
may lead to a phase separation between paired and unpaired components \cite{separation}.

Below we focus on the case of equal densities of the carriers on each nanotube when the system comprises of
three components:
the ground state of electron-hole pairs  (the superfluid  component), the quasiparticle excitations
above the ground state (the normal component) and the unpaired normal electron component on the outer nanotube.
The contribution to the conductivity due to the
unpaired component on the outer nanotube may be significant even at
zero temperature. Let us consider a finite amount of unpaired electrons on
the outer nanotube, with the conductivity
$\sigma^{un}_{ee}$.
We assume that  in the first approximation the unpaired electron component
moves with the same average velocity as the normal component \cite{halat}. This will result in
a greater current on the outer nanotube, than on the inner one. The difference will
correspond to the flow of the unpaired component.
In this case the effective electron
conductivity coefficient (the conductivity of the outer nanotube) can be estimated as
$\sigma^{eff}_{ee}=\sigma_{ee}+\sigma^{un}_{ee}$. The
conductivity for the unpaired component can be calculated as
$\sigma^{un}_{ee}=\frac{e^2\rho_{un} \tau}{m_e}$, where $m_e$ is the mass of an electron, and $\tau$
is the characteristic scattering time. The
density of the unpaired electrons is simply $\rho_{un}=\rho
\frac{R_{max}-R_{min}}{R_{max}}$. In the last expression $\rho$ is the concentration
of the carriers on both nanotubes and $R_{max}$, $R_{min}$ are the radii of the outer and inner 
nanotube, respectively.

Note that if the carriers density is not very high, the unpaired
electrons may strongly interact with Cooper pairs.
 In two coaxial nanotubes
in the dilute limit the electrons and holes form electron-hole bounded states-
indirect excitons \cite{12}. Moreover in the dilute limit if the numbers of electrons and holes are
different, in the quantum wells the carriers can even form bounded states
of trions \cite{13,14,15,trions}. In dilute limit the trions can be formed in the system of two coaxial
nanotubes as well. In this case if the outer nanotube is doped by
electrons we have to deal with negative trions $X^-$, while if it is doped by holes the
positively charged trions $X^+$ are formed.

In the case of unequal electron and hole densities resulting in
unbalanced chemical potentials, the system behavior becomes more
complicated. The BCS superfluid phase is energetically
unfavored, if the difference in the chemical potentials
exceeds  $\Delta_0/\sqrt{2}$, where $\Delta_0$ is the BCS gap
function at $T =0$ K and  $\mu_e=\mu_h$.  The system is expected to
form inhomogeneous superfluid phases, where the Cooper pairs
have non-zero total momentum. This is the case of the
Larkin-Ovchinnikov-Fulde-Ferrel 
phase, discussed, for example, in ~Refs.\cite{Fulde,Larkin,unbalanced}. In the present study we
assume equal chemical potentials for the electrons and holes that
also leads to close electron and hole densities.

\section{The electron-hole BCS state in two coaxial nanotubes}
\label{bdg}

 Let us consider a system of two thin cylindrical
nanotubes  of radii $R_{min}$ and $R_{max}$, and assume that the
nanotubes are doped in such way that the electrons are  the
carriers on the outer nanotube, and holes are the carriers on
the inner one. The effective Hamiltonian of this system can be written as \cite{Lozovik}
\begin{eqnarray}
\label{Heff}
H_{eff}=\sum_{\bf{p}}
(\xi_{\bf{p}}-\mu_e) b^\dagger_{\bf p}
b_{\bf p}+\sum_{\bf{p'}} (\xi'_{\bf{p'}}-\mu_h) a^\dagger_{\bf p'} a_{\bf p'} 
+ \nonumber \\ +\sum_{\bf{p},\bf{p'}} \left[\Delta_{\bf p, \bf{p'}} b^\dagger_{\bf p}  
a^\dagger_{-\bf p'}+H.c.\right],
\end{eqnarray}
where $a_{\bf p'}$ is the operator of annihilation of a hole
 on the inner nanotube, and $b_{\bf p}$
is the operator of annihilation of an electron on the outer
nanotube.
The single-particle
 eigenenergies of the electrons on the outer nanotube are given by
\begin{eqnarray}
\label{kin_el}
\xi_{\bf{p}}=\frac{ |{\bf p}|^2}{2 m_e}=
\frac{\hbar^2 }{2 m_e}\left[\frac{N^2}{d^2}+
\frac{m^2}{R_{max}^2}+k_z^2\right],
\end{eqnarray}
that correspond to the linear motion of an electron with the
momentum $p_z=\hbar k_z$ and the angular motion with the angular
momentum $L_e=R_{max}p_\phi=R_{max}\times \hbar\frac{m}{R_{max}}=\hbar m$,
$m=0,\pm 1, \pm 2,...$. Similarly the single-particle
 eigenenergies of the holes confined on the inner nanotube are  presented as
\begin{eqnarray}
\label{kin_hole}
\xi'_{\bf{p}'}=\frac{ {|{\bf p}'}|^2}{2 m_h}=
\frac{\hbar^2}{2 m_{h}}\left[\frac{N^2}{d^2}+\frac{m^2}{R_{min}^2}+k_z'^2\right].
\end{eqnarray}
 The angular momentum  of a hole is $L_h=R_{min} p'_\phi=R_{min}\times
\hbar\frac{m}{R_{min}}=\hbar m$, $m=0,\pm 1, \pm 2,...$. In Eqs.(\ref{kin_el}) and (\ref{kin_hole}) $m_e$
and  $m_h$ are the effective masses of electrons and holes and the
expression $\frac{\hbar^2 }{2 m_{e,h}} \frac{N^2}{d^2}$ in the
energies corresponds to the radial confinement of the particles. We
assume the identical thickness $d$ for  the both nanotubes, and that $d$ is sufficiently  smaller than the
radii of the nanotubes $d<<R_{min},R_{max}$. Within this
assumption  below we use only the lowest energy state for the radial
component of the eigenenergy $N=1$. Note the asymmetry between the
electron and hole excitation  spectra  exist in the system
due to different radii of the nanotubes, even for equal masses of the electrons and holes,  $m_e=m_h$.
The nonzero order parameter $\Delta_{\bf p, \bf{p'}}$ shows
that the system is in the superfluid phase. 
One has to keep in mind that it is assumed  the  electron-hole pairing
 occurs under the condition of the zero angular momentum of
the pair. It results in canceling  the linear momenta
components for the electron and hole $\hbar k_z+\hbar k'_z=0$, and
also canceling their angular momenta: $L_e+L_h=0$. The latter
condition leads immediately to $m+m'=0$.

Let's diagonalize Eq.(\ref{Heff}) using Bogoliubov unitary transformations
$a_{\bf{p}}=u_{\bf p} \alpha^\dagger_{-p}+v_{\bf p} \beta^\dagger_{-p}$ and
$b_{\bf{p}}=u_{\bf p} \beta_{\bf{p}}-v_{\bf p} \alpha_{\bf{p}}$,
 with the amplitudes $v_{\bf p}$ and $u_{\bf p}$. 
Following Ref.\cite{VM}, we assume that the order parameter
is independent on the momentum: $\Delta_{{\bf p },{\bf p'}} \equiv \Delta$.
Thus, the self-consistency condition for the order parameter has the form \cite{Lozovik}
\begin{eqnarray}
\label{order_param} \Delta=\frac{U}{S_{eff}}
\sum_{\bf{p}}  u_{\bf p} v_{\bf p}(1-f(E_{+})-f(E_{-})),    
\end{eqnarray}
where $U$ is the effective attractive interaction between
electrons and holes, and $S_{eff}=\pi L\sqrt{R_{min} R_{max}}$. 
Here we use the
notation $E_{\pm}=E_{\bf p}\pm\eta_{\bf p}$, $E_{\bf p}=\sqrt{\epsilon_{\bf p}^2+\Delta^2}$,
$\epsilon_{\bf p}=(\xi_{\bf{p}}+\xi'_{\bf{p}}-\mu_e-\mu_h)/2$,
$\eta_{\bf p}=(\xi_{\bf{p}}-\mu_e-\xi'_{\bf{p}}+\mu_h)/2$. In Eq.(\ref{order_param}) Fermi-Dirac
distribution function is given by $f(\epsilon)
=[\exp(\epsilon/(k_{B}T)) + 1]^{-1}$, where $k_{B}$ is Boltzmann constant, and $T$ is temperature. 
The amplitudes $u_{\bf p}$, $v_{\bf p}$ are given by
\begin{eqnarray}
u_{\bf{p}}^2=\frac{1}{2}(1+\frac{\epsilon_{\bf p}}{E_{\bf p}}), \nonumber \\
v_{\bf{p}}^2=\frac{1}{2}(1-\frac{\epsilon_{\bf p}}{E_{\bf p}}).
\end{eqnarray}
Note  that a dielectric with a high dielectric constant between the nanotubes
can significantly reduce the effective interaction $U$ that will dramatically
reduce the  BCS transition temperature $T_c$. The similar
effect is obtained in the case of  the increase of the separation distance
$D=R_{max}-R_{min}$ between the nanotubes. For two axial nanotubes with a relatively small
separation distance $D\ll R_{min},R_{max}$ on can use the result for the
interaction potential obtained for the electron-hole pairing in two
parallel plane layer, given by  Eq.(7) in Ref.\cite{Lozovik}. Assuming
the constant attraction 
of the paired particles one can use the simplified expression
$U=\frac{ e^2}{2\varepsilon \varepsilon_0 k_F}\exp(-D k_F)$, where  $k_F$  is the Fermi wavevector,
  $\varepsilon$ is the characteristic dielectric constant  of the dielectric between the coaxial nanotubes.
The cut-off in the sum given by Eq.~(\ref{order_param}) is set by the
characteristic plasma frequency \cite{Lozovik}.
The chemical potentials for electrons $\mu_e$ and holes $\mu_h$  are coupled
self-consistently to the  given surface densities of the
carriers $\rho_{e,h}=N_{e,h}/S_{e,h}$. $N_e$ and $N_h$ are the numbers of
 electrons and holes, respectively, while the surface of
the outer nanotube  is $S_e= 2\pi R_{max} L$, the surface
of the inner nanotube is $S_h=2\pi R_{min} L$, and $L$ is the length of each nanotube. Making the
statistical  averaging for the operators
$\hat{N}_h=\sum_{\bf{p}} a^\dagger_{\bf{p}} a_{\bf{p}}$ and
$\hat{N}_e=\sum_{\bf{p}} b^\dagger_{\bf{p}} b_{\bf{p}}$, one obtains
 \begin{eqnarray}
\label{dens}
\rho_e= \frac{2}{S_e} \sum_{\bf{p}}\left[ f(E_{+}) u^2_{\bf{p}}+
(1-f(E_{-}))v^2_{\bf{p}}\right], \nonumber\\
\rho_h= \frac{2}{S_h} \sum_{\bf{p}}\left[ f(E_{-}) u^2_{\bf{p}}+
(1-f(E_{+}))v^2_{\bf{p}}\right].
\end{eqnarray}
In our simulations we assume the electron surface density
$\rho_e$  is fixed on the outer nanotube and  the
 chemical potentials $\mu=\mu_e=\mu_h$  are equal.
Then $\mu$, $\Delta$ and $\rho_h$ are determined  by the
simultaneous solution of Eqs.~(\ref{order_param}) and (\ref{dens}). Since the
chemical potential and the order parameter mutually depend on each
other, one needs to run through several iterations. Note, for metals
the chemical potential is usually  larger by many orders of
magnitude  than the order parameter. This significantly
simplifies the calculations, because the increase of the order
parameter below the critical temperature practically does not affect
the chemical potential. So one can safely calculate the chemical
potential by setting the order parameter to zero, i.e. in the normal
state. For the system under consideration the order parameter value
can reach up to $10\%$ of the chemical potential, that requires the
full self-consistent solution of the problem. The calculations stop
when the convergence condition is met. The chemical potential and
the order parameter  has to satisfy the condition, that for
two consequent iterations the variations of their values are less
than $0.1\%$. It usually takes about several dozens of iterations to
converge.

\section{Calculation of the  conductivities }

\label{drag_coef}
For  the system of two coaxial nanotubes we introduce $\sigma_{ee}$ and $\sigma_{hh}$
as the longitudinal  quasiparticle conductivities for electron and hole nanotubes, and $\sigma_{he}$
and $\sigma_{eh}$  as the quasiparticle transconductivities.
The quasiparticle conductivities
can be obtained using  Gorkov-Nambu Green's function
formalism ~\cite{VM,Schrieffer,mahan} that gives the following expressions for the electron
\begin{eqnarray}
\label{con} \sigma_{ee} =-\frac{e^2}{S_e}\sum_{\bf{p}}\frac{\pi
{|\bf p}|^2}{m^2_{e}} \int_{-\infty}^{+\infty}\frac{\partial
f(\epsilon)}{\partial \epsilon} A^2_{ee}({\bf p},\epsilon)d \epsilon,
\end{eqnarray}
 and hole conductivities:
 \begin{eqnarray}
\label{con0} \sigma_{hh} =-\frac{e^2}{S_h}\sum_{\bf{p}'}\frac{\pi
{|\bf p}'|^2}{m^2_{h}} \int_{-\infty}^{+\infty}\frac{\partial
f(\epsilon)}{\partial \epsilon} A^2_{hh}({\bf p}',\epsilon)d \epsilon.
\end{eqnarray}
Note, that the summation over the momenta ${\bf p}$ and ${\bf p}'$ are not equivalent,
because of the different quantization of the angular part for the inner and outer nanotubes.
 The transconductivities are obtained in a similar way:
\begin{eqnarray}
\label{con1} \sigma_{ij} =\frac{e^2}{S_i} \sum_{\bf{p}, {\bf p}'}\frac{\pi
{\bf p}{\bf p}'}{m_{i}m_{j}}\int_{-\infty}^{+\infty}\frac{\partial
f(\epsilon)}{\partial \epsilon} A^2_{ij}({\bf p},\epsilon)d \epsilon,
\end{eqnarray}
where $(i,j) = (e,h)$ are the indices corresponding to the outer or
inner nanotubes and the summation is done over the momenta with matching quantum numbers. 
Note that the momenta ${\bf p}$ and ${\bf p}'$  contain the
continuous  linear $p_z, p'_z$ and the discrete angular $p_\phi,p'_\phi$ components in Eq.~(\ref{con1}).

The matrix spectral function $\hat{A}({\bf{p}} ,\epsilon)$ is given by 
\begin{eqnarray}
\label{aij} \hat{A}({\bf{p}} ,\omega) = - \frac{1}{\pi} \mathrm{Im}
{\hat{G}}({\bf{p}} , \epsilon + i \delta),
\end{eqnarray}
where $\hat{G}$ is the  Gorkov-Nambu matrix Green's
function, and $\delta$ is an infinitesimal positive energy.
In the absence of disorder we have ${\hat{G}} =
{\hat{G}}^{0}$, and ${\hat{G}}^{0}$ determines the unperturbed spectral functions:
\begin{eqnarray}
\label{g0} 
A^0_{ee}({\bf p},\epsilon)=v_{\bf p}^2\delta(\epsilon-E_{+})+u_{\bf p}^2\delta(\epsilon+E_{-}),  \nonumber \\
A^0_{hh}({\bf p},\epsilon)=u_{\bf p}^2\delta(\epsilon-E_{+})+v_{\bf p}^2\delta(\epsilon+E_{-}), \nonumber \\
A^0_{eh}({\bf p},\epsilon)=\left(\frac{\Delta}{2 E_{\bf p}}\right)(\delta(\epsilon-E_{+})-\delta(\epsilon+E_{-})). 
\end{eqnarray}

In the presence of  weak impurities the spectral functions become\cite{mahan}:
\begin{eqnarray}
\label{pert} 
A_{ee}({\bf p},\epsilon)=v_{\bf p}^4 t_h \delta(\epsilon-E_{+})+u_{\bf p}^4 t_e \delta(\epsilon+E_{-}),  \nonumber \\
A_{hh}({\bf p},\epsilon)=u_{\bf p}^4 t_h \delta(\epsilon-E_{+})+v_{\bf p}^4 t_e \delta(\epsilon+E_{-}), \nonumber \\
A_{eh}({\bf p},\epsilon)=\left(\frac{\Delta}{2 E_{\bf p}}\right)(t_h \delta(\epsilon-E_{+})+t_e \delta(\epsilon+E_{-})). 
\end{eqnarray}

In the presence of impurities 
 the conductivity for the outer nanotube $\sigma_{ee}$ can be
calculated as
\begin{eqnarray}
\label{sigma_ee} \sigma_{ee} =-\frac{e^2 \pi}{S_e} \sum_{p_z,p_\phi}\frac{
p_z^2+p_\phi^2}{m^2_{e} k_B T} \times \nonumber \\
\left[ v_{\bf p}^4  \frac{t_h}{\cosh^2\left(\frac{E_{+}}{2 k_BT}\right)} + 
u_{\bf p}^4 \frac{t_e}{{\cosh^2\left(\frac{E_{-}}{2 k_B T}\right)}}\right].
\end{eqnarray}
To evaluate this expression, the $p_z$ was quantized
using $p_z=l \hbar/L$ where $l=0,\pm1,\pm2,...$. 
    In the simulations the length of the nanotube $L$ is 
    set to be relatively big to simulate an infinite system, particularly $L=1000$ nm.
Similarly one can represent the conductivity for the inner nanotube
$\sigma_{hh}$:
\begin{eqnarray}
\label{sigma_hh} \sigma_{hh}=-\frac{e^2 \pi}{S_h} \sum_{p'_z,p'_\phi}\frac{
{p'}_z^2+p_\phi'^2}{m^2_{h} k_B T}\times \nonumber \\
\left[ u_{\bf p}^4 \frac{t_h}{\cosh^2\left(\frac{E_{+}}{2 k_BT}\right)} + 
v_{\bf p}^4 \frac{t_e}{{\cosh^2\left(\frac{E_{-}}{2 k_B T}\right)}}\right].
\end{eqnarray}
 The interlayer transconductivity $\sigma_{eh}$ can be similarly expressed as
\begin{eqnarray}
\label{sigma_eh} \sigma_{eh} =\frac{e^2 \pi}{ S_e}
\sum_{p_z,p_z',p_\phi,p'_\phi}\frac{p_z {p'}_z+p_{\phi}p'_{\phi}
}{m_{e} m_h k_B T} \times \nonumber \\ 
\left[ \frac{t_h}{\cosh^2\left(\frac{E_{+}}{2 k_BT}\right)} 
+ \frac{t_e}{{\cosh^2\left(\frac{E_{-}}{2 k_B T}\right)}}\right]
u_{\bf p}^2 v_{\bf p}^2.
\end{eqnarray}
The corresponding transconductivity $\sigma_{he}=\sigma_{eh}\frac{R_{max}}{R_{min}}$ because in the
derivation of the Kubo's formula~\onlinecite{mahan} the induced
current should be averaged over the area where it is flowing.
Therefore, since the transconductivity $\sigma_{eh}$ describes the
current of electrons on the outer nanotube induced by the holes
current on the inner nanotube, the normalization area should be
$S_e$, that is reflected in   Eq.~(\ref{con1}). Note the asymmetry in the
transconductivities $\sigma_{eh}\ne\sigma_{he}$ exists only for the
spatially separated electrons and holes in two colaxial  nanotubes
due to the difference between their radii, while the relation
$\sigma_{eh}= \sigma_{he}$ holds for two parallel plane
layers~\cite{VM}.
Follow  Ref.~\onlinecite{VM} the scattering lifetimes of
quasiparticles can be evaluated as:
\begin{eqnarray}
\label{tau0}
t_{h}^{-1} &=& t_{hn}^{-1} u_{\bf p}^4 \left|\frac{\partial E_{+}}{\partial \xi'_{\bf p}}\right|^{-1}+ t_{en}^{-1}v_{\bf p}^4 \left|\frac{\partial E_{+}}{\partial \xi_{\bf p}}\right|^{-1} \nonumber\\
t_{e}^{-1} &=& t_{hn}^{-1} v_{\bf p}^4 \left|\frac{\partial E_{-}}{\partial \xi'_{\bf p}}\right|^{-1}+ t_{en}^{-1}u_{\bf p}^4 \left|\frac{\partial E_{-}}{\partial \xi_{\bf p}}\right|^{-1}.
\end{eqnarray}
In Eq.~(\ref{tau0}) $t_{en}$ and $t_{hn}$ are the electron and hole scattering times
in the normal state.
The  electron and hole scattering times $t_{en}$ and $t_{hn}$ are related to the hole $V^h_{{\bf p} {\bf p'}}$
and electron $V^e_{{\bf p} {\bf p'}}$ impurities scattering potentials 
as follows:
\begin{eqnarray}
\label{taun}
t_{hn}^{-1} = 4\pi \sum_{\bf p'} <|V^h_{{\bf p} {\bf p'}}|^2> \delta(\xi'_{\bf p}-\xi'_{\bf p'}), \nonumber\\
t_{en}^{-1} = 4\pi \sum_{\bf p'} <|V^e_{{\bf p} {\bf p'}}|^2> \delta(\xi_{\bf p}-\xi_{\bf p'}),
\end{eqnarray}
where $<..>$ denotes the statistical average over the inpurities. It is assumed that the impurieties on the inner and 
outer nanotubes are uncorrelated: $<V^e V^h>=0$.
Using the chain rule
one obtains: $|\frac{\partial E_{+}}{\partial
\xi_{\bf p}}|=|\frac{\epsilon_{\bf p}}{E_{\bf p}}y+y_1|$, $|\frac{\partial
E_{-}}{\partial \xi_{\bf p}}|=|\frac{\epsilon_{\bf p}}{E_{\bf p}}y-y_1|$,
$|\frac{\partial E_{+}}{\partial
\xi_{\bf p}'}|=|\frac{\epsilon_{\bf p}}{E_{\bf p}}y_2+y_3|$ and
$|\frac{\partial E_{-}}{\partial
\xi_{\bf p}'}|=|\frac{\epsilon_{\bf p}}{E_{\bf p}}y_2-y_3|$.
Here we used a notation $y=\frac{1}{4}(2+x+xw)$,
$y_1=\frac{1}{4}(x+xw-2)$,
$y_2=\frac{1}{4}(2+x^{-1}+(xw)^{-1})$,
$y_3=\frac{1}{4}(x^{-1}+(xw)^{-1}-2)$,
where $x=\frac{m_e}{m_h}$ and $w=\frac{R^2_{max}}{R^2_{min}}$.
Note that for $R_{min}=R_{max}$ Eq.~(\ref{tau0}) reduces to
the expression given by Eq.~16 in  Ref.\onlinecite{VM}. Since we are interested to 
study the effects of the asymmetry between electrons and holes due to
different radii of the nanotubes, in our
simulations we assume the  electrons and holes  have equal masses:
$m_e=m_h=m$. We also assume that the scattering times of electrons and holes in the normal state are equal: 
$t_{en}=t_{hn}=t_n$. Since the electron and hole 
scattering times enter the conductivities as linear weighting coefficients, we 
do not expect any qualitative change of our results in the case of unequal scattering times. 
Similar conclusion was expressed
also in Ref. ~\onlinecite{VM}, where different electron and hole masses and lifetimes were used.  
 Let us emphasize that the transconductivities $\sigma_{eh}$
and $\sigma_{he}$ are negative~\cite{VM}, because for a system of
spatially separated charges the Coulomb drag induces  the
currents  flowing in the opposite directions.


In general, the currents are carried by the BCS superfluid and
quasiparticles \cite{VM}. We characterize the quasiparticle contributions in
the linear-response regime as follows:
\begin{eqnarray}
\label{drag}
\left(\begin{array}{c} { j}_{h}\\ { j}_{e}
\end{array}\right)=\left(\begin{array}{cc} \sigma_{hh} & \sigma_{he}\\
\sigma_{eh}&\sigma_{ee}
\end{array}\right)\left(\begin{array}{c} { E}_{h}\\ { E}_{e}
\end{array}\right),
\end{eqnarray}
where ${ j}_{e(h)}$ and ${ E}_{e(h)}$ are the current
flows of quasiparticles  and electric field in electron (hole)
nanotubes, correspondingly.

Let us mention that the electric fields in electron and hole
nanotubes are identical~\cite{VM}:
${E} \equiv {E}_{e} = {E}_{h}$. Therefore from
Eq.~(\ref{drag}) we obtain
\begin{eqnarray}
\label{ef} E = \frac{j_{e} + j_{h}}{\sigma_{ee}
+\sigma_{hh} + \sigma_{he} + \sigma_{eh}} \ .
\end{eqnarray}
Let us consider the drag setup, where a fixed current $j_{0}$ flows
through the drive nanotube and the voltage drop across the second
nanotube, namely the drag-nanotube, is measured, $j_{h0} =j_{0}$ and
$j_{e0}=0$. Using Eq.~(\ref{ef}) for $j_{e} + j_{h} = j_{0}$, the
voltage in the drag nanotube $V_{drag}$ can be written as
\begin{eqnarray}
\label{vol} V_{drag} = L E = \frac{L j_{0}}{\sigma_{ee}+ \sigma_{hh} + \sigma_{he} + \sigma_{eh}}.
\end{eqnarray}


\begin{figure} 
\includegraphics[width=8cm,angle=0]{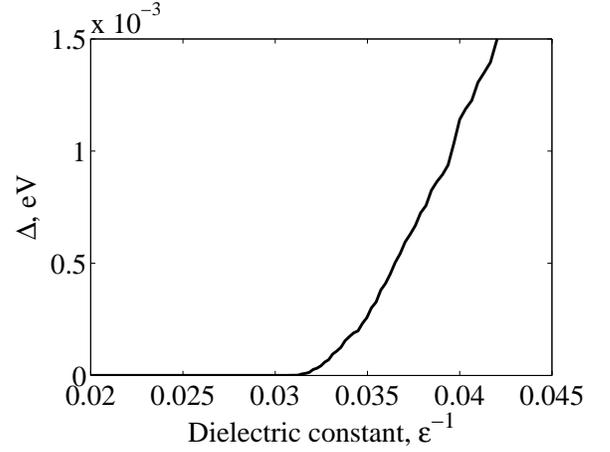}
\caption{{\protect\footnotesize {The  order parameter
 $\Delta$  as a function of the  inverse dielectric constant $\varepsilon^{-1}$.
The radii of the nanotubes are  $R_{min}=6$~nm and $R_{max}=8$~nm.
The surface carrier concentration is $\rho=10^{12}$~cm$^{-2}$
and the temperature is $T=0$~K.
 }}}
\label{label_figure2}
\end{figure}

\begin{figure} 
\includegraphics[width=8cm,angle=0]{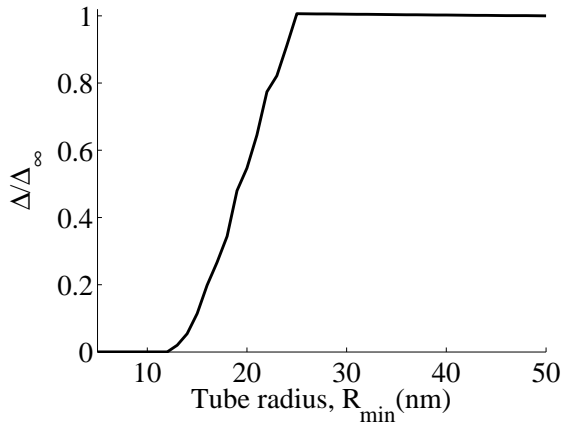}
\caption{{\protect\footnotesize {The normalized order parameter
 $\Delta$  as function of the  radius of the inner nanotube $R_{min}$.
The radii difference  $D=R_{max}-R_{min}$ is kept constant $D=2$~nm.
The surface carrier concentration is $\rho=10^{12}$~cm$^{-2}$
and the temperature is $T=0$~K.
The order parameter is normalized to its value $\Delta_{\infty}$ for the flat layers geometry, that corresponds to 
$R_{min},R_{max}\to\infty$.
 }}}
\label{label_figure3}
\end{figure}

\begin{figure} 
\includegraphics[width=8cm,angle=0]{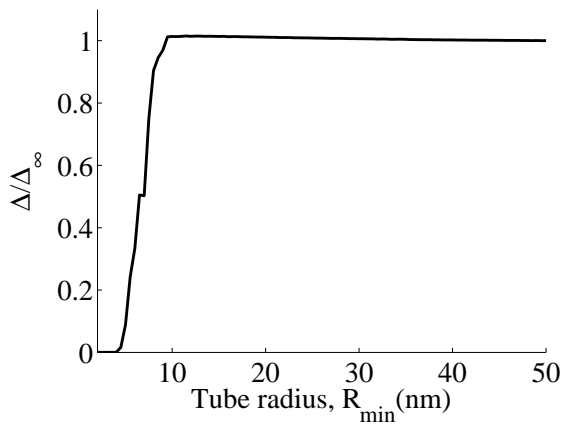}
\caption{{\protect\footnotesize {The normalized order parameter
 $\Delta$  as function of the  radius of the inner nanotube $R_{min}$.
 Note the electron-hole interaction is $50\%$
 stronger than in Fig.~\ref{label_figure3}.
The radii difference  $D=R_{max}-R_{min}$ is kept constant: $D=2$~nm.
The surface carrier concentration is $\rho=10^{12}$~cm$^{-2}$
and the temperature $T=0$~K.
The order parameter is normalized to its value $\Delta_{\infty}$ for the flat layers geometry, that corresponds to 
$R_{min},R_{max}\to\infty$.
 }}}
\label{label_figure4}
\end{figure}

\begin{figure} 
\includegraphics[width=8cm,angle=0]{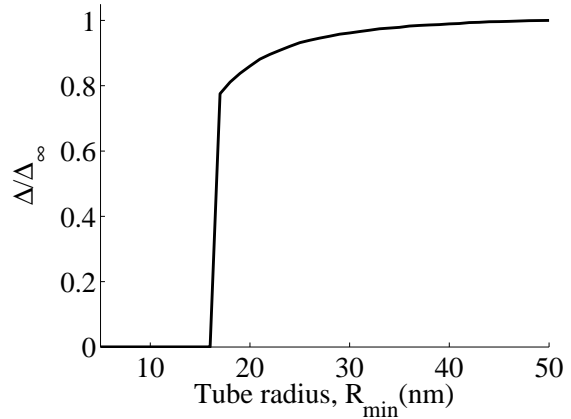}
\caption{{\protect\footnotesize {The normalized order parameter
 $\Delta$  as a function of the  radius of the inner nanotube $R_{min}$.
 The electron-hole interaction is the same as in Fig.~\ref{label_figure4}.
The radii difference  $D=R_{max}-R_{min}$ is kept constant $D=2$~nm.
The surface carrier concentration is $\rho=10^{12}$~cm$^{-2}$
and the temperature is  $T=0.5 T_c$, where $T_c$ is the critical temperature for the flat layers geometry.
The order parameter is normalized to its value $\Delta_{\infty}$
for the flat layers geometry, that corresponds to 
$R_{min},R_{max}\to\infty$.
 }}}
\label{label_figure5}
\end{figure}

\begin{figure} 
\includegraphics[width=8cm,angle=0]{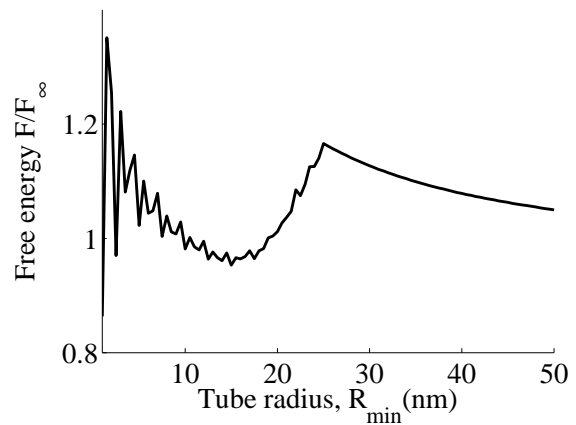}
\caption{{\protect\footnotesize {The normalized free energy density
as function of the  radius of the inner nanotube $R_{min}$.
The radii difference  $D=R_{max}-R_{min}$ is kept constant $D=2$~nm.
The surface carrier concentration is $\rho=10^{12}$~cm$^{-2}$
and the temperature is $T=0$~K.
The free energy is normalized to its value $F_{\infty}$ for the flat layers geometry, that corresponds to 
$R_{min},R_{max}\to\infty$.
 }}}
\label{label_figure8}
\end{figure}

\begin{figure} 
\includegraphics[width=8cm,angle=0]{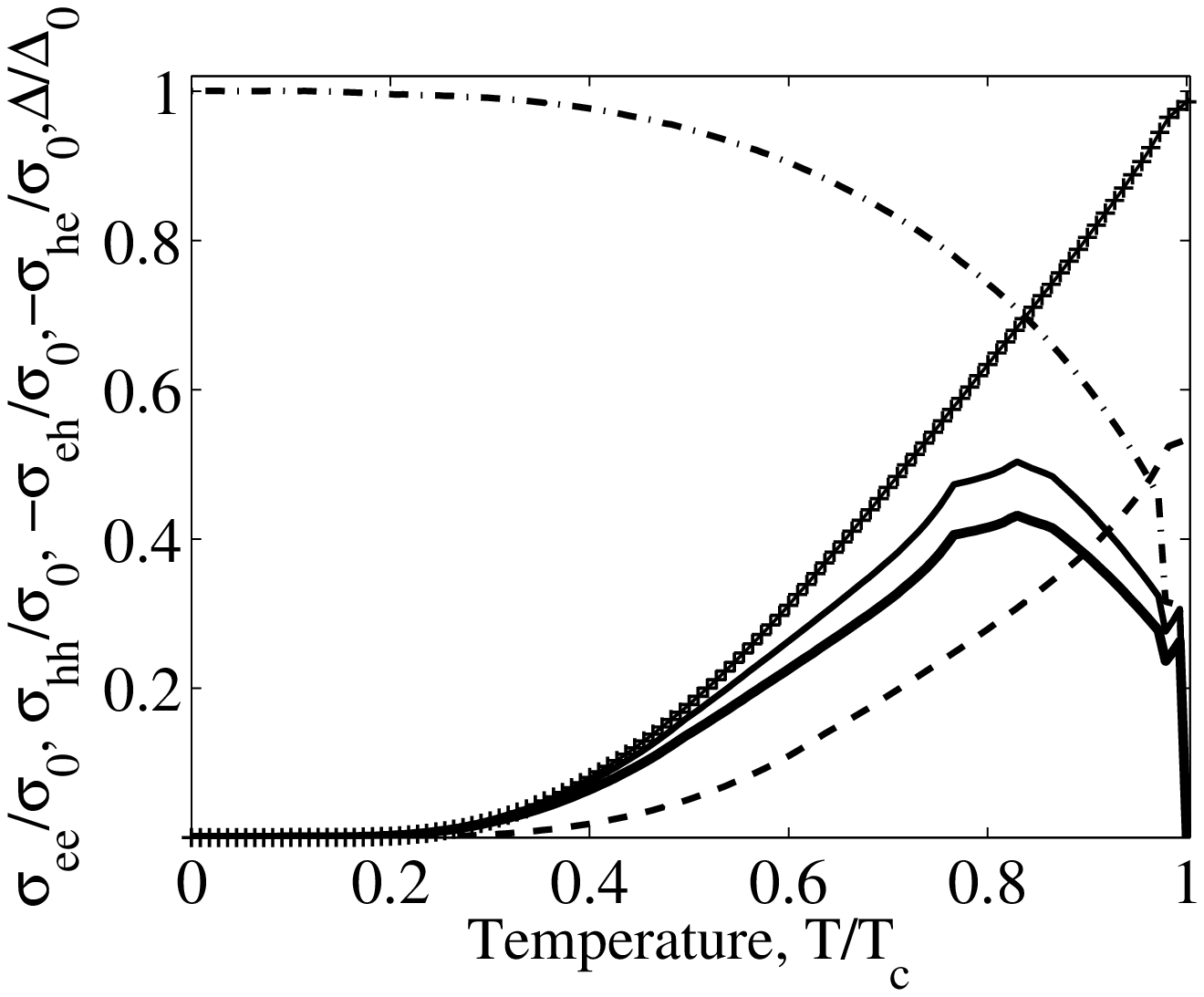}
\caption{{\protect\footnotesize { The normalized quasiparticles
conductivities $\sigma_{ee}$ (line with the plus markers),
$\sigma_{hh}$ (dashed line), $\sigma_{eh}$ (thin solid line),
$\sigma_{he}$ (thick solid line) and the order parameter $\Delta$
(dash-dotted line) as functions of the normalized temperature
$T/T_c$. The radii of the nanotubes are $R_{min}=5$~nm and
$R_{max}=7$~nm,
 the surface carrier concentration is $\rho=10^{12}$~cm$^{-2}$.
 Note the conductivities are normalized
to the electron conductivity of the outer nanotube $\sigma_{0}$ at the normal state,
and the order parameter is normalized to its value $\Delta_0$ at $T=0$~K.}
 }}
\label{label_figure6}
\end{figure}

\begin{figure} 
\includegraphics[width=8cm,angle=0]{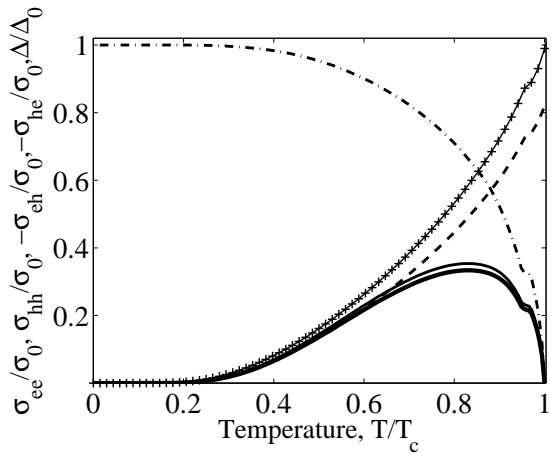}
\caption{{\protect\footnotesize {The normalized quasiparticles
conductivities $\sigma_{ee}$ (line with the plus markers),
$\sigma_{hh}$ (dashed line), $\sigma_{eh}$ (thin solid line),
$\sigma_{he}$ (thick solid line) and the order parameter $\Delta$
(dash-dotted line) as functions of the normalized temperature
$T/T_c$. The radii of the nanotubes are $R_{min}=50$~nm and
$R_{max}=52$~nm,
 the surface carrier concentration is $\rho=10^{12}$~cm$^{-2}$.
 Note the conductivities are normalized
to the electron conductivity of the outer nanotube $\sigma_{0}$ at the normal state,
and the order parameter is normalized to the order parameter $\Delta_0$ at $T=0$~K.
$\sigma_{he}$ and $\sigma_{eh}$ are so close that they are indistinguishable on this figure.}
 }}
\label{label_figure7}
\end{figure}

\section{Discussion and conclusions}

\label{disc}
 For all presented calculations the surface electron
concentration is set to $10^{12}$~cm$^{-2}$. We have also performed calculations using much higher density,
$10^{14}$~cm$^{-2}$. We have found that the increase of the density does not lead to qualitative changes of
the presented in this work results.
The hole concentration is obtained self-consistently
and under the condition of equal chemical potentials it is very close to the concentration of electrons.

First, we study the role of electron-hole attraction in the system on the formation of the superfluid phase.
It is known that in  the case of the asymmetry  caused  by
the difference between the masses of the electrons and holes there
is  the critical strength of the electron-hole attraction,
under which there is no BCS transition \cite{asym}. In a system of
two coaxial  nanotubes the strength of interaction can be
controlled, for example, by changing the dielectric constant of
the dielectric between the nanotubes. To illustrate this we calculate the dependence of the order
parameter on the inverse dielectric constant presented in Fig.~\ref{label_figure2}. From  Fig.~\ref{label_figure2}
one can see that there is  a critical value for the
interaction strength. For a weaker electron-hole interaction strength
(or higher value of the dielectric constant) there is no BCS state.
In our calculations the radii of the nanotubes are set to
$R_{min}$=6~nm and $R_{max}$=8~nm and the temperature $T=0$~K.
Note that for different radii of the
nanotubes the asymmetry between electrons and holes will be
different, that will correspond to different critical values of the
interaction strength. In  general, for a  larger asymmetry
one would expect a higher critical value  of the interaction
strength. For the symmetric case of two parallel plane layers the 
change in attraction strength between the carriers results in the change of the gap, which follows the standard 
BCS theory.

Second, we investigate how relatively large electron-hole asymmetry
can destroy the superfluid phase in the system. The results of the calculations
for the dependence of the normalized order parameter on the radius of the inner nanotube
for the different strength of the electron-hole interaction and for different temperature are presented
in Figs.~\ref{label_figure3} - \ref{label_figure5}.
According to the results of our calculations, the BCS state
still can  be formed in the case of  moderate
asymmetry between the electrons and holes excitation spectra. By
decreasing the radii $R_{min}$ and $R_{max}$ of the nanotubes while
keeping the constant separation distance $D=R_{max}-R_{min}$ between the nanotubes one can
monotonously increase the asymmetry in the system. In the
simulations shown in Figs.~\ref{label_figure3},~\ref{label_figure4} and
\ref{label_figure5} the separation distance is set to $D=2$~nm and the dielectric constant between the nanotubes is
set to $\varepsilon=27$.
As it seen from Fig.~\ref{label_figure3} for
relatively large radii ($R_{min}$,~$R_{max}>25$~nm) the order
parameter in the system converges to its value for the plane layers
system. At zero temperature $T=0$~K,  as the
radii decrease, the order parameter has a kink at $\approx 25$ nm and then
almost linearly drops to zero for the inner radius of the nanotube
$12$~nm$<R_{min}<25$~nm. One can suggest that for a fixed attraction
strength between electrons and holes there is a critical asymmetry, and
that for a higher degree of the asymmetry there is no BCS
transition. The nature of the kink in Fig.~\ref{label_figure3}  has to be studied in
more details.  The oscillations in the dependence of the order
parameter and critical temperature on the size of superconducting
nanowires~\cite{Shanenko_2006,Shanenko_2007}, superconducting
films~\cite{Shanenko_2007}, superconducting metallic
grains~\cite{Altshuler} were obtained in the framework of BCS and
Bogoliubov--de Gennes theories caused by quantum size effect. The
experimental demonstration of the quantum size effects in the
dependence of the critical temperature on the size of the
superconducting film  was observed in Ref.~\onlinecite{Guo}. The
similar kink to the one obtained in our Paper for the ground state
energy respect to the coupling constant was treated as a first-order
quantum phase transition in the orthogonal-dimer spin
chain.~\cite{Koga}. The possibility of the first order quantum phase
transition resulting in the point of non-analyticity in the ground
energy as a function of the nanotube radius will be studied in the
future.

To further investigate the effect of the electron-hole asymmetry in Fig.~\ref{label_figure8} we show the normalized free energy density in the system as a function
of the inner radius. At zero temperature the free energy is equal to the ground state energy of the system. 
For the calculations the difference $R_{max}-R_{min}=2$~nm is kept constant. 
The dielectric constant and temperature are the same as for Fig.~\ref{label_figure3}.
The free energy density has also a kink at the radius $R_{min}\approx25$~nm, which is the same radius, at which 
the kink in the  order   parameter is observed in Fig.~\ref{label_figure3}. This should not be a 
surprise, since the ground state
of a BCS superconductor contains a term proportional to $\Delta^2$ \cite{Schrieffer}.
The oscillations in the dependence of the ground state
energy on the nanotube radius $R_{min}$ for $R_{min}<25$~nm  seems to be caused 
by the quantum size effects.

For a stronger attraction between the electrons and holes, which is
modeled by a dielectric with $1.5$ times smaller dielectric
constant $\varepsilon=18$, the system  can sustain a higher degree of the
asymmetry that is shown in Fig.~\ref{label_figure4}. At  $T=0$~K the order
parameter increases essentially with the decrease of the radii
of the nanotubes. This effect is attributed to the quantum
confinement and it was predicted for homogeneous \cite{Shanenko} and
inhomogeneous
 \cite{Ilya} superconducting nanowires. For  the radii less than the critical radii
 $R_{min}\approx 10$~nm the order parameter again shows an almost linear drop to zero.

The similar drop can be observed at the finite temperature
$T=0.5 T_c$ presented in Fig.~\ref{label_figure5}, where $T_c$ is the critical temperature
 for the plane layers system. 
 In Fig.~\ref{label_figure5} the critical radius
 $R_{min}\approx 17$~nm is considerably bigger than the critical radius for $T=0$~K. 
 One can also notice that the
 enhancement of the order parameter due to the quantum confinement does not
 take place at this sufficiently high temperature.

Finally, we study the temperature dependence of the transconductivity
coefficients in the system. We have found there are non-zero drag conductivities
$\sigma_{eh}\ne\sigma_{he}$  between the two coaxial nanotubes at
temperatures below the critical temperature of the BCS phase
transition $T_{c}$. Since the drag conductivity is a signature of
the BCS state, the measurements of the drag conductivity can be used
to demonstrate the existence of the BCS state resulting in the
electron-hole superfluidity in the system and superconductivity in
each nanotube. The measurement of the temperature corresponding to
vanishing of the drag conductivity can be treated as the measurement
of $T_{c}$. Note in the limiting case of low carrier concentrations
the exciton pairs will be formed Ref.\onlinecite{12}, that will result in
a finite drag  even at high temperatures due to the Coulomb
attraction and formation of bounded states for electron-hole pairs.
The results of calculation of the order
parameter and the normalized conductivity coefficients as functions
of the temperature are presented in Figs.~\ref{label_figure6} and \ref{label_figure7}.
The conductivity coefficients are normalized to
the conductivity of the outer nanotube in the normal state $\sigma_{en}$,
and, therefore, they do not depend on the  scattering time of the
carriers, which is assumed equal for  electrons and holes.

To understand the dependence of the conductivities and the order parameter on temperature we considered
two scenarios and perform calculations for two systems when the radii of the nanotubes
are different by the order of magnitude, however the separation distance between the inner and outer cylinders was kept constant:
$D=2$~nm. Particularly, we considered the system when
 the inner nanotube has $R_{min}=5$~nm and the
outer one  has  $R_{min}=7$~nm and  a larger system with
$R_{min}=50$~nm and $R_{min}=52$~nm. The results of calculations are
presented in Figs.~\ref{label_figure6} and ~\ref{label_figure7},
correspondingly. Note that we plot the conductivities in the units of the  conductivity of the outer nanotube 
$\sigma_{en}$  in the normal state. 
The comparison of the results of the calculations
demonstrates that as the temperature $T$ approaches the critical
temperature of the phase transition $T_{c}$,
 the transconductivities $\sigma_{eh}$ and 
 $\sigma_{he}$ vanish, because the quasiparticle drag conductivity appears
 due to the existence of
the electron-hole pairing caused by the Coulomb attraction, and when the order parameter $\Delta$
 vanishes, the conductivities  $\sigma_{eh}$ and $\sigma_{he}$  also vanish.
At low temperatures the conductivities are
exponentially suppressed. The similar behavior  was obtained for
an infinite two plane layer
  electron-hole system in  Ref.~\onlinecite{VM}.
  We also would like to emphasize that $\sigma_{ee}\ne \sigma_{hh}$ due  to the difference between the
  radii of the nanotubes, and the quasiparticle transconductivity $\sigma_{eh}$ takes values
of the same order of magnitude as $\sigma_{ee}$ and $\sigma_{hh}$.
 The difference between the transconductivities $\sigma_{eh}$
and $\sigma_{he}$ is also the specific property of the spatially
separated electrons and holes in two coaxial nanotubes, while for
two plane electron and hole layers one has
$\sigma_{eh}=\sigma_{he}$~\cite{VM}. Note that in Figs.~\ref{label_figure6} and ~\ref{label_figure7} the
normalized gap function has little smooth kinks at some temperatures. This corresponds to 
the mutual effects between the order parameter and the chemical potential in the system.
Unlike in superconducting metals, in our calculations the ratio  $\Delta/\mu$ is  of the order of $0.1$, so the changes of the gap
with the temperature affects the chemical potential and through it the conductivity.

To summarize,  a system of two coaxial
nanotubes with  spatially separated electrons and holes on each
nanotube was studied theoretically.  It was shown that at the temperatures lower than
the critical temperature the system undergoes BCS-like transition.
The transition results in correlations between the carriers on each
nanotube and as a result, nonzero Coulomb drag coefficient.
The measurement of  the drag coefficient can be used  to monitor the
BCS transition in the system. We analyzed how the asymmetry
between electrons and holes controlled through the radii of the
nanotubes, affects the formation of the BCS state in the system. We
found that there is a kink in the dependence of the order parameter and the ground state energy
on the asymmetry with the monotonous change of the asymmetry in the
system. For relatively stronger attraction between the electrons and
holes and sufficiently low temperatures the system can exhibit the
enhancement of the order parameter due to the quantum confinement
effect.  We also anticipate for a system of two coaxial nanotubes with
equal carriers concentrations the contribution to the conductivity
due to the unpaired component on the outer nanotube.
The considered system has an intrinsic asymmetry due to
different radii of the nanotubes. One can utilize this asymmetry in nanoscale
devices. Unlike the ratio of the transconductivities $\sigma_{eh}/\sigma_{he}$, 
for relatively small radii of the  nanotubes the ratio $\sigma_{ee}/\sigma_{hh}$ is 
not equal to the ratio of radii, 
and therefore, the ratio of the amount of the carriers on the nanotubes.  
Therefore,  a situation is possible where 
the inner drive nanotube can induce a higher current
in the outer drag nanotube. In other words, the drag current will be higher than the drive one. 
The potential drop on the outer nanotube will be lower than on the inner one.
Therefore, the system  functions as a step down transformer at zero frequency.
In the case of using the outer nanotube as the drive, the system works as a step up transformer, where 
the drag current will be lower than the drive current.
In the case of ideal drag the transformation coefficient is given by the ratio of the radii of the nanotubes.
One can possibly manufacture such systems using multiwall
carbon nanotubes.

\acknowledgments

The authors are grateful to G. Vignale for valuable discussions.
O.L.B. acknowledges the support of KITP, Santa Barbara under National Science 
Foundation Grant No. PHY-1125915, and their kind hospitality at the early stages of this project.

\end{document}